\documentclass{elsart}

\usepackage{amssymb}

\usepackage{epsfig,float,wrapfig,amsmath,graphicx}
\usepackage[hang,small,ruled,bf]{caption}

\newcommand{\be}{\begin{equation}}
\newcommand{\ee}{\end{equation}}
\newcommand{\beq}{\begin{eqnarray}}
\newcommand{\eeq}{\end{eqnarray}}

\begin{document}

\begin{frontmatter}

\title{Subthreshold  \mbox {\boldmath $\phi$} meson production 
in heavy-ion collisions}

\author[Dresden]{H.W. Barz},
\ead{H.W.Barz@fz-rossendorf.de}
\author[Budapest]{M. Z\'et\'enyi},
\ead{zetenyi@rmki.kfki.hu}
\author[Budapest]{Gy. Wolf},
\ead{wolf@rmki.kfki.hu}
\author[Dresden]{B. K\"ampfer}
\ead{kaempfer@fz-rossendorf.de}

\address[Dresden]{Forschungszentrum Rossendorf, 
Pf 510119, 01314 Dresden, Germany}
\address[Budapest]{KFKI Research Institute for Particle and Nuclear Physics,
POB 49, H-1525 Budapest, Hungary}

\begin{abstract}
Within a transport code of BUU type the production of $\phi$
mesons in the reactions Ni + Ni at 1.93 A$\cdot$GeV
and Ru + Ru at 1.69 A$\cdot$GeV is studied.
New elementary reaction channels 
$\rho$N($\Delta) \to \phi$N and $\pi$N(1520) $\to \phi$N
are included.
In spite of a substantial increase of the $\phi$ multiplicities
by these channels the results stay below the
tentative numbers extracted from experimental data.  
\end{abstract}

\end{frontmatter}

\section{Introduction}

Meson production in heavy-ion collisions is considered
a sensitive probe of the reaction dynamics \cite{ref_1}.
Particularly
interesting is the production of heavy mesons 
at such a kinetic energy which is insufficient
to create the respective meson in a free, single nucleon-nucleon 
collision.
Secondary and cooperative processes are needed to accumulate
the energy for the creation of the heavy meson.
Therefore, the study of the subthreshold particle production
is thought to reveal many aspects of the behaviour of strongly
interacting matter including the 
equation of state \cite{Aichelin_Ko}.

Focusing on heavy-ion collisions at bombarding energies in the
range of one A$\cdot$GeV the production of strangeness degrees of freedom
attracted much attention, both experimentally and theoretically.
For instance, the
recent measurements seem to comply with a noticeable 
mass splitting between the $K^-$ and $K^+$
mesons \cite{ref_2}, as predicted by various effective (chiral) models
for pseudoscalar mesons \cite{ref_3}.

In contrast to the comparatively rich body of experimental data
on open strangeness meson production, the 
information on mesons with hidden strange\-ness, i.e. the  $\phi$,
is scarce.
Only a few measurements
\cite{herrmann,kotte} for the reactions Ni + Ni at 1.93 A$\cdot$GeV
and Ru + Ru at 1.69 A$\cdot$GeV have been reported. 
In these measurements a strongly limited amount of the phase 
space was accessible. Extrapolations to the full
phase space point to a surprisingly large production cross section.
The results of the analysis \cite{chung2} for the
reaction Ni + Ni \cite{herrmann} let one recognise that the calculations 
based on a transport model underestimate the data by more than
a factor of ten. 
(Notice that refined re-analyses \cite{kotte,FOPImeeting} 
of this reaction has shifted the
previously reported $\phi$ yield \cite{herrmann} to a larger value.) 
In the calculations \cite{chung2} the $\phi$ mesons are solely
produced in the elementary processes 
BB $\to$ $\phi$BB and $\pi$B $\to$ $\phi$B, where the symbol
 B denotes the baryons N and $\Delta$.
Very recently the analysis of the $\phi$ production within
another transport model has been repeated  \cite{hartnack}
with the same production channels confirming the previous finding
\cite{chung2}. 
The discrepancy of these calculations with the data 
\cite{kotte,FOPImeeting,kottenew} has inspired us to 
study whether further elementary channels might essentially
contribute to the $\phi$ meson production. 

In Ref.~\cite{barz1}
the role of three-nucleon collisions for
$\phi$ production was investigated.
It turned out that these processes are well described 
by a sequence of two two-body processes inside the dense medium.
However, to obtain the correct yield, 
one has to include all intermediate particles which
can be created inside the medium. This concerns 
especially the $\rho$ mesons which have a rather large 
production cross section
and a low threshold energy when colliding  with baryons.

Furthermore, even at subthreshold bombarding
energies of about 1.5 up to 2 GeV per nucleon,
up to 30\% of the nucleons
are excited to $\Delta$ and N$^*$ resonances 
which have a lower threshold for
$\phi$ meson production than ground state nucleons when colliding  
with other particles. 
It is the aim of this study to investigate the effects of these
additional production channels which have not been studied so far.
To be specific, we consider here the elementary reactions
$\rho$B $\to$ B$\phi$ and $\pi$N(1520) $\to$ N$\phi$.

The study of the $\phi$ meson production is also of interest
concerning the $K^-$ production. 
Due to the strong decay channel
$\phi \to K^+ K^-$ a substantial feeding of
the $K^-$ channel could not be ruled out. 
This is particularly tempting as the
analysis \cite{kotte} points to a substantial fraction of $K^-$
stemming from $\phi$ decays.
In the near future also the
decay channel $\phi \to e^+ e^-$
becomes experimentally accessible with the HADES detector
at the heavy-ion synchrotron SIS at GSI/Darmstadt \cite{HADES}.
Due to the large $s\bar{s}$ quark content, $\phi$ mesons
are expected to interact much weaker with the non-strange
baryonic matter in accordance with the Okubo-Zweig-Iizuka rule
(cf.\ \cite{Ellis}).
Nevertheless, noticeable in-medium effects, e.g.\
mass shift and broadening, are
expected at higher temperature and density
\cite{hatsuda,weise}.

Our paper is organised as follows. In section 2 the 
cross sections of the above mentioned
elementary hadron reactions are estimated within a one-boson
exchange model. Sections 3 and 4  describe the resulting $\phi$
multiplicities when implementing these new reaction channels in a
transport code. In section 5 we compare with 
available experimental results.
The summary and the discussion can be found in section 6.

\section{Elementary cross sections} 

We are going to describe heavy-ion collisions within the 
Boltzmann-Uehling-Uhlenbeck (BUU) formalism. At beam energies around
2 GeV per nucleon many resonances are generated. 
Therefore, there are many possibilities to produce $\phi$ mesons 
by two-body collisions. In Refs.~\cite{chung2,hartnack} the cross sections
of the collisions of the most abundant particles, 
NN $\to$ NN $\phi$, N$\Delta$ $\to$ NN$\phi$,  
$\Delta \Delta \to$ NN $\phi$, $\pi$N $\to$ N$\phi$, 
and $\pi\Delta\to$ N$\phi$, 
have been used. Since the cross sections are not well known 
a one-boson exchange model has been employed to estimate them
\cite{chung2,titov2}. 
The parameters of the corresponding Lagrangians have been 
fixed by decay widths and cross sections measured at different energies.
Selected new data near threshold became recently available \cite{Jim}.
As mentioned above,
employing these cross sections within transport
codes one gets $\phi$ meson production rates
which seriously underestimate the extrapolations made from the measurements
\cite{herrmann,kotte}. 

Channels with $\rho$ mesons as
incoming particles have not been included in Refs.~\cite{chung2,hartnack}. 
Since the $\phi$ meson 
decays into the non-strange sector overwhelmingly 
via the $\pi \rho$ channel (with 13\% branching ratio, see \cite{PDG}), 
both mesons are expected to play an important role for the 
$\phi$ production. 
Other channels are substantially weaker. 

\subsection{$\phi$ meson production in $\rho$-baryon collisions}

\begin{figure}
\begin{center}
\includegraphics[width=8.6cm]{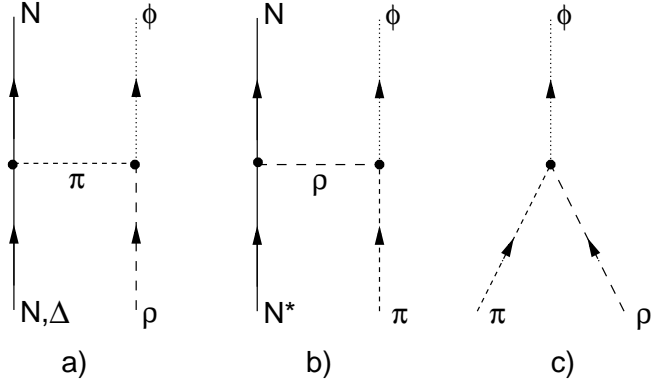}
\end{center}
\caption{Diagrammatic representation of processes contributing to the
reactions (a)
$\rho$ (N,$\Delta$) $\to$ $\phi$ N, (b)
$\pi$ N$^*$ $\to$ $\phi$ N and (c) $\pi \rho$ $\to$ $\phi$.}
\label{diagrams}
\end{figure}

Let us consider first the $\phi$ production in $\rho$ induced
reactions on the nucleon and the $\Delta$, i.e.
$\rho N \to \phi N$ and $\rho\Delta \to \phi N$.
Figure \ref{diagrams}a shows the corresponding 
tree level diagram hitherto not yet considered.
We calculate the cross sections within the one-boson
exchange model with the effective interaction Lagrangians
\beq
{\mathcal L}_{\pi NN} & = & \frac{f_{\pi NN}}{m_{\pi}}
\bar{\psi}\gamma_5\gamma^{\mu}\vec{\tau}\psi\cdot\partial_{\mu}\vec{\pi}, \\
{\mathcal L}_{\pi N\Delta} & = & - \frac{f_{\pi N\Delta}}{m_{\pi}}
\bar{\psi}\vec{T}\psi^{\mu}\cdot\partial_{\mu}\vec{\pi} \;+\; \mbox{h.c.}, \\
{\mathcal L}_{\pi\rho\phi} & = & \frac{f_{\pi\rho\phi}}{m_{\phi}}
\epsilon_{\mu\nu\alpha\beta}
\partial^{\mu}\phi^{\nu}\partial^{\alpha}\vec{\rho}^{\beta}\cdot\vec{\pi}.
\eeq
Here $\psi$ is the nucleon field, $\psi_{\mu}$ is the Rarita-Schwinger field 
for the $\Delta$ resonance, and $\pi$, $\vec{\rho_{\mu}}$, 
$\phi_{\mu}$ are the meson fields while
$m_{\pi}$ and $m_{\phi}$ stand for the pion and $\rho$ meson masses, 
respectively. The symbol $\vec{\tau}$ denotes the isospin Pauli matrix,
 and $\vec{T}$ is the isospin $1/2 \to 3/2$ transition matrix.
The diagrams in Fig.~\ref{diagrams} are to be understood
as guidance to construct Lorentz invariant amplitudes from
effective interaction Lagrangians. They must be supplemented
by additional elements to achieve a reasonable
description of elementary hadron processes, 
for which data are available. In this line of arguments   
we introduce multipole form factors for off-shell pions 
with four-momentum $p_\pi$ at each vertex
which, e.g.\ for the $\pi NN$ vertex reads
\begin{equation}
F^\pi_{\pi NN}(p_\pi^2) = \Bigg( 
\frac{(\Lambda^\pi_{\pi NN})^2 - m_\pi^2}
{(\Lambda^\pi_{\pi NN})^2 - p_\pi^2} \Bigg)^n
\end{equation}
with $n=1$, while for $\rho$ mesons we use 
a corresponding expression with $n=2$.
The cross section for the $\rho\Delta \to \phi N$
reaction is regularised by the Peierls method \cite{Peierls}.
We adopted the coupling constants $f_i$ and cutoff parameters $\Lambda_i$
for the 
$\pi$ NN and $\pi$N$\Delta$ vertices from the  Model II in Table B.1 
of Ref.~\cite{machleidt}. For $f_{\pi\rho\phi}$ and 
$\Lambda^{\pi}_{\pi\rho\phi}$ we used the values given in \cite{chung2}.

The isospin averaged
differential cross sections can be written as
\begin{equation}
\frac{d\sigma}{d\Omega} =
\frac{1}{64\pi^2 s}
\frac{\vert{\bf p}_{\rm out,CM}\vert}{\vert{\bf p}_{\rm in,CM}\vert}
\vert {\mathcal M} \vert^2,
\label{totalxsec}
\end{equation}
with $s$ being the square of the center-of-mass energy and  
${\bf p}_{\rm in (out), CM}$ the incoming (outgoing) three-momentum 
in the center-of-mass
system.
The quantity $\vert {\mathcal M} \vert^2$ 
represents the squared matrix elements summed over the final and averaged 
over the initial spin and isospin states. These quantities are given by
\begin{equation}
\vert {\mathcal M}_{\rho N \to \phi N}\vert^2 =
\frac{16}{3}
\left(\frac{A_{\rho N \to \phi N}}{p_{\pi}^2-m_{\pi}^2}\right)^2 m_N^2
\left(p_N \cdot p_{N^{\prime}} - m_N^2 \right)
\left( (p_{\rho}\cdot p_{\phi})^2 - m_{\rho}^2 m_{\phi}^2 \right)
\end{equation}
for the $\rho$N  channel and
\begin{multline}
\vert {\mathcal M}_{\rho \Delta \to \phi N}\vert^2 =
 \frac{4}{27}
\left(\frac{A_{\rho \Delta \to \phi N}}{p_{\pi}^2-m_{\pi}^2}\right)^2 \\
\times
\left(p_{\Delta}\cdot p_N + m_{\Delta}m_N \right)
\left( \left(\frac{p_{\Delta}\cdot p_N}{m_\Delta}\right)^2 -  m_N^2 \right)
\left( (p_{\rho}\cdot p_{\phi})^2 - m_{\rho}^2 m_{\phi}^2 \right)
\end{multline}
for the $\rho \Delta$ channel.
Here $p_N$, $p_{\Delta}$, $p_{\rho}$, and $p_{\phi}$ denote the
four-momenta of the respective particles. In the case of the $\rho$N
channel, $p_N$ and $p_N^{\prime}$ 
refer to the four-momenta of the incoming
and outgoing nucleon, respectively.
The factors $A_i$ are given by $A_{\rho N \to \phi N} = 
(f_{\pi N N}f_{\pi\rho\phi}/m_{\pi}m_{\phi})
F^{\pi}_{\pi N N}(p_{\pi}^2)F^{\pi}_{\pi\rho\phi}(p_{\pi}^2)$ and
$A_{\rho \Delta \to \phi N} = 
(f_{\pi N \Delta}f_{\pi\rho\phi}/m_{\pi}m_{\phi})
F^{\pi}_{\pi N \Delta}(p_{\pi}^2)F^{\pi}_{\pi\rho\phi}(p_{\pi}^2)$.
The total cross sections are obtained by numerically integrating 
Eq.~(\ref{totalxsec}) over the solid angle $\Omega$.

Cross sections of the specific isospin channels can be deduced
from the isospin
averaged cross section by multiplying with an appropriate isospin factor.
These isospin factors, which can be expressed in terms of Clebsch-Gordan
coefficients, are given in Table \ref{isofac}.

\begin{table}[t]
\begin{center}
\begin{tabular}{c|cc|cccc}
 & n & p & $\Delta^-$ & $\Delta^0$ & $\Delta^+$ & $\Delta^{++}$ \\
\hline
$\rho^-$ & 0 & 2 & 0 & 0 & 1 & 3\\
$\rho^0$ & 1 & 1 & 0 & 2 & 2 & 0\\
$\rho^+$ & 2 & 0 & 3 & 1 & 0 & 0\\
\end{tabular}
\end{center}
\caption{Isospin factors for the various channels of the reactions
$\rho$N $\to \phi$N and $\rho \Delta \to \phi$N.}
\label{isofac}
\end{table}

\subsection{$\phi$ formation from the $\pi \rho$ channel}

The cross section of the $\pi \rho $ process has a 
very narrow resonance structure with a width of 
$\Gamma_{tot} =$ 4.43 MeV and a branching ratio of
$\Gamma_{\pi\rho}/\Gamma_{tot}$ = 0.13 in the $\pi\rho$ channel. The
formation cross section, depicted in Fig.~\ref{diagrams}c,  
of the $\phi$ meson 
in the $\pi^+ \rho^-$ or the $\pi^- \rho^+$ channel is correspondingly
\beq
        \sigma = \frac{ \pi}{2{\bf p}_{in}^2} 
         \frac{\Gamma_{\pi\rho}\Gamma_{tot}}
         {(\sqrt{s}-m_\phi)^2+\Gamma_{tot}^2/4}
  \label{BWigner1}
\eeq
which gives a maximum cross section of 9.5 mb at resonance
peak position. The $\pi^0 \rho^0$ channel is isospin forbidden.

\subsection{Effects of higher resonances} 

At bombarding energies of about 2 GeV per nucleon baryons can be 
highly excited. Consequently the production threshold
is reduced by their excitation energy if $\phi$ mesons are created
in collisions with other particles.
To estimate the importance
of such processes we test the influence
of the collision of a $\pi$ meson  with the N(1520) 3/2$^-$ resonance. 
The corresponding diagram is  exhibited in 
Fig.~\ref{diagrams}b. This
resonance is chosen because it decays with a large branching ratio
of about (20 $\pm$ 5)\% \cite{PDG,langg} into the $\rho$ channel,
although this channel would be closed for the nominal masses of the $\rho$ 
and N(1520). This hints to the fact 
that a large part of the production cross section
proceeds via virtual $\rho$ propagation as described by the diagram
in Fig.~\ref{diagrams}b.
This assumption may not be applicable to higher resonances
which can emit $\rho$ mesons nearly on shell.

There are several possibilities to describe the 
$\rho$NN(1520) vertex by using simple effective interaction 
Lagrangians:
\beq
\label{sca1520}
{\mathcal L}_{\rho NN(1520)} & = & g_0
\bar{\psi}_N \vec{\tau} \psi_{1520}^\mu \,\vec{\rho}_\mu \;+\;\mbox{h.c.},\\
\label{vec1520}
{\mathcal L}_{\rho NN(1520)} & = & \frac{f_1}{m_\rho}
\bar{\psi}_N \vec{\tau} \gamma ^\nu \psi_{1520}^\mu \,
(\partial_\nu \vec{\rho}_\mu - \partial_\mu \vec{\rho}_\nu) 
\;+\;\mbox{h.c.},\\
\label{ten1520}
{\mathcal L}_{\rho NN(1520)} & = & \frac{f_2}{m_\rho^2}
\bar{\psi}_N \vec{\tau} \sigma^{\alpha\beta} \psi_{1520}^\mu \,
\partial_\alpha \partial_\mu \vec{\rho}_\beta  \;+\;\mbox{h.c.}\;.
\eeq
The second Lagrangian is related to the vector dominance approach
(see e.g. \cite{peters}) while the last one follows from 
constituent-quark models \cite{riska}.
We fix the coupling constants by assuming that each Lagrangian 
alone would provide 
the partial width to the $\rho$(two-pion) 
decay channel. Such a procedure supplies upper limits on the
coupling constants.
We obtain the values $g_0$ = 7.5, $f_1$ = 10.5 
and $f_2$ = 68, respectively. 
With these values the Lagrangians (\ref{sca1520}) and 
(\ref{vec1520}) give roughly the same 
cross section near threshold. Our coefficient  $f_1$ is larger than
the corresponding coefficient in Ref.~\cite{peters}, which is caused
by the monopole form factor used there. 
In Fig.~\ref{crosssec1} we present the 
cross sections obtained with the Lagrangians (\ref{vec1520}) and 
(\ref{ten1520}).  The large coupling constant $f_2$ needed for
the Lagrangian (\ref{ten1520}) leads to a much larger 
cross section for the $\phi$ production than the 
Lagrangian (\ref{vec1520}).
This value of $f_2$, 
however, differs from  the value estimated 
from the quark model \cite{riska}. 
As our Lagrangians are nothing more than a covariant parametrisation
of the interaction  the coupling constants are generally
energy dependent. The larger relevant energy in $\phi$ production
compared to the N(1520) decay could be the reason for the different
estimated values of the coupling constants. In the following 
we use the cross section from the
Lagrangian (\ref{vec1520}) as standard in our BUU model calculations. 
We also performed calculations using the cross section calculated with
the Lagrangian (\ref{ten1520})
in order to estimate the uncertainties of the $\phi$ meson yield
from the $\pi$ + N(1520) channel, see Fig.~\ref{crosssec1}.

\subsection{Summary of cross sections} 

\begin{figure}
\begin{center}
\includegraphics[width=8.6cm]{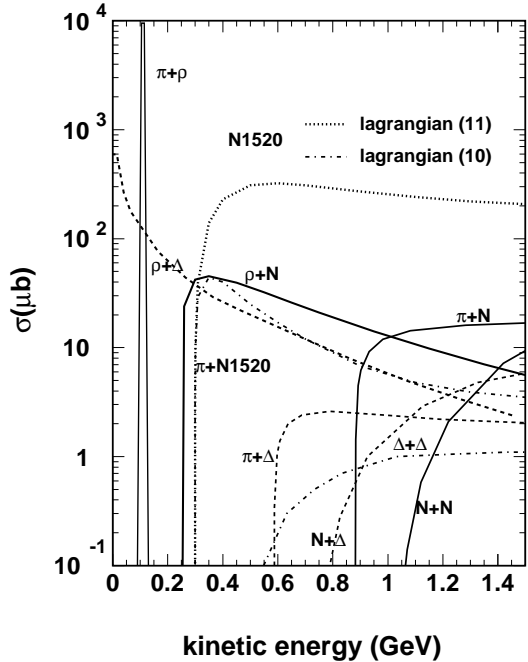}
\end{center}
\caption{Production cross sections for $\phi$ mesons for
several binary reactions as a function of the 
relative kinetic centre-of-mass energy in the entrance channel.
The cross sections are calculated for the nominal masses of
the resonances. Especially cross sections  for $\rho \Delta$ collisions
having different masses can considerably deviate from the curve shown.
}
\label{crosssec1}
\end{figure}

Summarising the results of this section we show the 
calculated individual cross sections in  
Fig.~\ref{crosssec1} as a function of the kinetic energy 
of the relative motion in the entrance channel.
Obviously, the $\rho$ B channels are very important at low energies.
We also studied the reaction $\pi \pi \to \phi \pi$ which, 
however, is not important
because of the large threshold energy which can rarely be
overcome by the pions. For the $\phi$ production via the BB and $\pi$B
channels we use the same cross sections as found in Ref.~\cite{chung2}.

\section{BUU calculations} 

To calculate $\phi$ meson  production in a heavy-ion reaction we
use the version of the BUU code introduced in Ref.~\cite{Wolf1}.
The production of $\phi$ mesons proceeds via the reactions
discussed in the previous section. To increase the statistics 
the perturbative method is used, i.e.\ if in a two-particle collision the 
threshold is overcome, the $\phi$ meson is created and weighted with
its production probability.
However the  production process
itself and its subsequent collisions do not influence the motion of 
the other participating particles. 
To simplify the calculations
an isotropic cross section is assumed for each 
elementary hadron reaction 
and in the case of a three particle
final state the energy distribution is taken from the phase space.

In the transport model all nucleon and delta resonances are included up to 
2.2 GeV. Furthermore, $\Lambda$ and $\Sigma$ baryons as well as 
the mesons $\pi, \eta, \rho, \sigma, \omega$, $K^{+}$ and $K^{0}$
are explicitly propagated.
The properties (masses, widths, branching ratios) of the baryon
resonances have been fitted \cite{Wolf2} 
to the one- and two-meson production data available
from  pion-nucleon reactions. The production cross sections
for resonances in nucleon-nucleon collision are derived from data
on meson production in nucleon-nucleon reactions.

Furthermore, momentum dependent potentials for the nucleons are
included. The same potentials are also used  for the $\Delta$ particles.
As standard potential 
we employ such one which delivers a soft equation of state with
an incompressibility parameter of 215 MeV.
Less is known about in-medium properties of
the $\phi$ mesons. While in Ref.~\cite{weise} a very small
mass shift has been found,
we use for exploratory purposes the prediction 
by Hatsuda et al. \cite{hatsuda} 
for the density dependence of the $\phi$ meson mass:
\beq
     m_\phi^*\,=\,m_\phi - 0.025 \,  m_\phi  \frac{n}{n_0}.
     \label{mhatsuda}
\eeq
The second term is treated as the mean field of the $\phi$ meson.

\begin{figure}[t]
\begin{center}
\includegraphics[width=8.6cm]{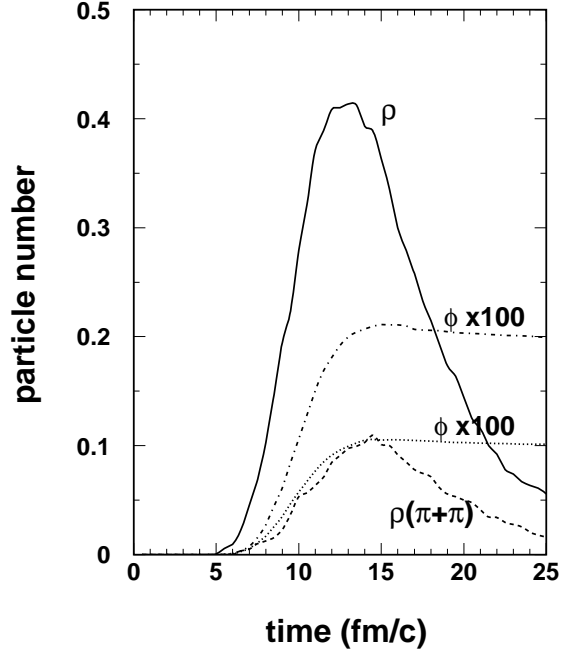}
\end{center}
\caption{Number of produced $\rho$ and $\phi$ mesons as a
function of time for the reaction Ru on Ru at
beam energy of 1.69  A$\cdot$GeV and impact parameter $b = 2.8$ fm. 
The lower pair of curves is calculated by switching off the 
$\rho$ production due to resonance decays. 
}
\label{phitime}
\end{figure}

The interaction of the $\phi$ mesons with surrounding 
nucleons is relatively weak. The 
elastic cross section is estimated to be in the order of 0.5 mb,
however, a strong absorption cross section is expected via the
reaction channel $\phi$B $\to $K$\Lambda$. Following Ref.~\cite{haglin}
we use a value of 
$6 \mbox{mb} \cdot \exp (-3.3\epsilon/\mbox{GeV})$  
with $\epsilon$ being
the kinetic energy in the entrance channel. 
Similar values are employed in 
Ref.~\cite{chung2}. 

Using these potentials we have calculated the $\phi$ production cross
sections for the reactions Ni on Ni at 1.93 A$\cdot$GeV and
Ru on Ru at 1.69 A$\cdot$GeV bombarding energy. The results are
reported in Table \ref{4piyield}.

\begin{table}[t]
\begin{center}
\begin{tabular}{l|c|c}
 yields from & Ni + Ni (1.93 GeV) & Ru + Ru (1.69 GeV) \\ 
\hline
   B + B           &  $3.5\cdot 10^{-4}$ &  $3.1 \cdot 10^{-4}$ \\
   $\pi$ + B       &  $2.9\cdot 10^{-4}$ &  $3.2 \cdot 10^{-4}$ \\
   $\rho$ + B      &  $8.9\cdot 10^{-4}$ &  $11.8\cdot 10^{-4}$ \\
   $\pi$ + $\rho$  &  $1.6\cdot 10^{-4}$ &  $1.5 \cdot 10^{-4}$ \\
   $\pi$ + N(1520) &  $0.5\cdot 10^{-4}$ &  $0.6 \cdot 10^{-4}$ \\
\hline
 total yield       &  $1.7\cdot 10^{-3}$ &  $2.0 \cdot 10^{-3}$ \\
 in HELITRON       &  $1.2\cdot 10^{-5}$ &  $1.5 \cdot 10^{-5}$ \\
 in CDC            &  $2.7\cdot 10^{-5}$ &  $2.9 \cdot 10^{-5}$ \\
\hline
experiment \cite{kotte} &  $(8.7\pm3.6)\cdot 10^{-3}$  &  
$(6.4\pm2.5)\cdot 10^{-3}$
\end{tabular}
\end{center}

\caption{Multiplicities of $\phi$ mesons 
per central event in Ni + Ni and Ru + Ru reactions. 
The first 5 lines give the results from the special channels indicated.
The symbol B comprises the nucleon and the $\Delta$ particle.
Lines 7 and 8 are the yields observed in the two detector subsystems.}
\label{4piyield}
\end{table}

\section{Influence of model parameters}

In the first five lines of Table \ref{4piyield} the contributions 
of the different production
channels are listed. We notice a substantial increase by a 
factor of 2.5 of the
$\phi$ multiplicities when including the new $\rho$ channels.
The $\pi$ + N(1520) reaction channel calculated with
the Lagrangian (\ref{vec1520}) does not essentially contribute
to the $\phi$ production. However, when using the cross sections
obtained from the Lagrangian (\ref{ten1520}) 
the $\phi$ meson yield from the
$\pi$ + N(1520) channel rises by more than an order of magnitude
to 6.6$\cdot$10$^{-4}$ (7.7$\cdot$10$^{-4}$) for the reaction
Ni + Ni (Ru + Ru). This indicates that the $\pi$ + N(1520) 
channel might not be negligible for the $\phi$ production,
but to clarify this requires a better knowledge 
of the $\rho$NN(1520) vertex.

In order to test the dependence of the $\phi$ meson
yield on the equation of state we carried out also calculations with
different potentials. For instance, employing a hard 
momentum dependent potential
with an incompressibility
parameter of \mbox{380 MeV} only two third of the $\phi$ 
yield of the soft 
equation of state has been obtained for the Ru on Ru reaction.
The reduction of the $\phi$ multiplicity concerns especially
the BB and $\pi$B channels while the $\rho$B and 
$\pi$$\rho$ channels are less influenced.
Obviously, to overcome the larger potentials
the collision partners use up a large amount of kinetic
energy which then 
lacks for the $\phi$ production. Similar effects are known 
for kaon production, see Ref.~\cite{ref_2}.
We have also found that momentum independent potentials
enhance the number of the produced $\phi$ mesons.

\begin{figure}
\begin{center}
\includegraphics[width=8.6cm]{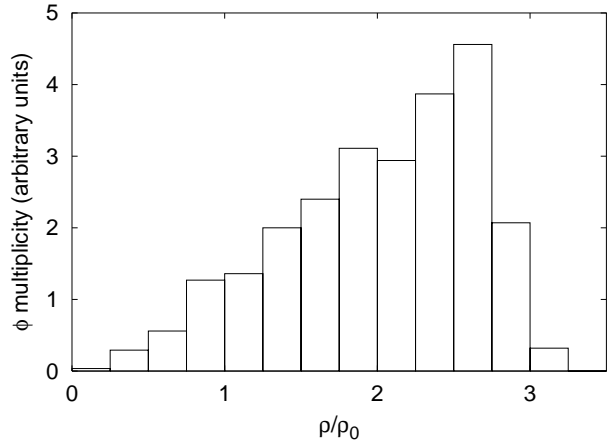}
\end{center}
\caption{Relative number of produced $\phi$ mesons as a function of the 
density at creation point for the reaction 
Ru + Ru at 1.69 A$\cdot$GeV beam energy}
\label{rhodep}
\end{figure}

The time dependence of the number of produced $\rho$ and $\phi$ mesons 
is exhibited in Fig.~\ref{phitime}. The time origin is chosen such  that
at 5 fm/c the nuclei touch. 
The maximum $\rho$ number is obtained at complete overlap (13 fm/c).
However, the maximum density
is attained very early at 9 fm/c. At this time
the $\phi$ production reaches its maximum value and ceases 
at 15 fm/c. 

In the calculations it turned out that indeed most of the 
$\rho$ mesons are generated by resonance decays; 
a substantially smaller part 
stems from the $\pi \pi$ annihilation process. 
To demonstrate the importance of the 
baryon resonance decay into $\rho$ mesons for
$\phi$ meson production we tentatively switched off
these decay channels. Then the $\rho$ number is diminished to 25\%.
This in turn reduces drastically the number of produced $\phi$'s
by a factor of two.

To estimate the importance of the in-medium effects of the $\phi$ mesons
in dense nuclear matter
it is interesting to see at which densities the $\phi$'s
are created. In Fig.~\ref{rhodep} we display the number of $\phi$ mesons
as a function of the local density at the creation point. 
The production reaches its maximum at around 2.5 times normal
nuclear density while the average density is about twice normal
nuclear matter density.
Nevertheless the effect of the mean field in Eq.~(\ref{mhatsuda})
is not dramatic. The yield increases by about 10\% only.
The effect is stronger in the BB channels.
Furthermore, the slope parameters of the momentum 
distribution (see discussion
in the following section) are reduced by 
about 5 MeV. One should be aware that the in-medium effect of $\phi$
mesons has to be treated together with the effects of the K mesons.
If e.g. in addition the sum of the K$^-$ and K$^+$ masses 
is reduced by 200 MeV (see \cite{ref_3}) 
the decay time of the  $\phi$ meson is
considerably shortened and could be smaller than 5 fm/c \cite{weise}.
Such an effect would decrease the detection probability
due to the occurrence of kaon rescattering. This effect
is not treated here.
Due to the $\phi$B$\to$K$\Lambda$ cross section about 30\% of the 
created $\phi$ mesons are reabsorbed.

In Ref.~\cite{chung2} a $\phi$ multiplicity of 3$\cdot$10$^{-4}$
has been obtained within a transport model treatment
for Ni on Ni collisions. This value is 
smaller than ours, even
if we do not count the contributions from the $\rho$
and N(1520) channels. Besides the use of different mean fields, 
one has to observe that in our calculations the effect of the
increasing decay widths due to the dropping K$^-$ mass has not 
been included. The $\phi$ formation from the K$^+$K$^-$ channel
is negligible as was already shown in Ref.~\cite{chung2}.

\section{Comparison with experiment} 

The FOPI collaboration has measured $\phi$ meson production
in the reactions Ni + Ni at 1.93 A$\cdot$GeV and 
Ru + Ru at 1.69 A$\cdot$GeV 
\cite{herrmann,kotte} via the decay channel $\phi \to K^+ K^-$.
The detector system allows to identify the original $\phi$'s 
in two separated parts of the momentum space.
The analysis \cite{herrmann} based on data
of identified K$^{\pm}$ pairs within the acceptance of the central drift
chamber (CDC) at target rapidity was recently supplemented 
\cite{kotte,kottenew} 
by taking into account also
the HELITRON forward Plastic Wall subsystem of
the FOPI detector which covers a rather limited phase space
near midrapidity.
A full phase-space extrapolation of the observed K$^{\pm}$ pairs
from the very limited acceptance regions of both detector systems
can be performed only by relying on assumptions on the
spectral shape of the $\phi$ meson distribution.
Using the information from both detector systems, 
CDC and HELITRON, and assuming  
a thermal distribution of the $\phi$
mesons, $dN/d{\bf p} \propto \exp{(-E/T_{\rm eff})}$, 
one has been  able to estimate 
the total yields given in the last line of Table \ref{4piyield}.
Thereby a rather low temperature
of \mbox{$T_{\rm eff}$ = 59 (72) MeV}
has been  obtained for the reaction Ni + Ni (Ru + Ru). 

Our calculations have been carried out  for
central events  comprising about 9\% (15\%) of the total
cross section in the symmetric collisions
of Ni (Ru) nuclei.
Despite the gain of $\phi$ mesons due to the large 
contribution of the $\rho$ mesons the calculated numbers are still
smaller than the $\phi$ multiplicities extrapolated from experiment.
For completeness we list also in Table 2 the $\phi$ yields
within the acceptance regions of the CDC and HELITRON subdetectors.

In our calculations we find that the momentum spectrum of the 
$\phi$ mesons is not compatible with a thermal distribution.
The distribution is stretched out in
longitudinal direction. We can attribute
two different slope parameters $T_\perp$ and $T_\parallel$ for the
perpendicular and longitudinal distribution, respectively.
Fitting independently the distributions near midrapidity  
and at small transverse momenta 
we find for the Ru + Ru (Ni + Ni) collision 
$T_\perp$ = 90 (100) MeV and $T_\parallel$ = 113 (120) MeV, 
respectively. 
The rescattering of $\phi$ mesons increases the longitudinal
temperature by about 10 MeV while the transverse temperature
is less affected. 

\begin{figure}
\begin{center}
\includegraphics[width=8.6cm]{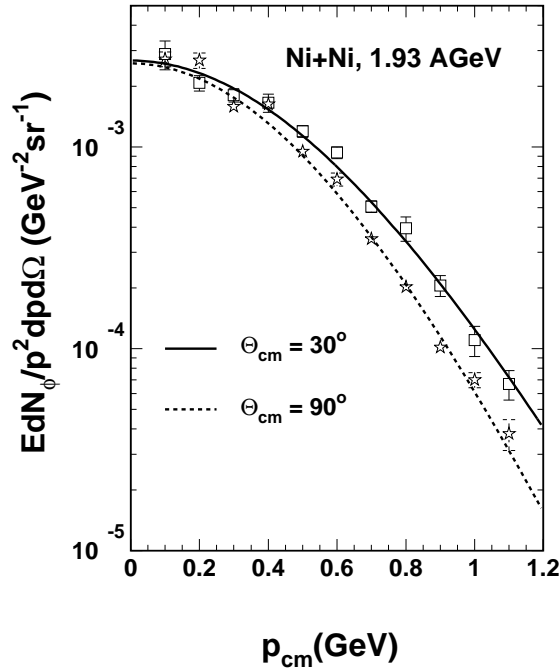}
\end{center}
\caption{Predicted momentum distributions of $\phi$ mesons 
in the centre-of-mass system for central  Ni + Ni collisions
for two centre-of-mass angles $\Theta$. 
The error bars on the symbols reflect the statistical uncertainty
in our Monte-Carlo calculation. 
The smooth
curves show fits to a thermal model with temperature values 
of 120 MeV for $\Theta_{cm}$ = 30$^o$ and 100 MeV for 
$\Theta_{cm}$ = 90$^o$, respectively. }
\label{siginv}
\end{figure}

To illustrate the effect of the anisotropy 
of the $\phi$ meson emission we exhibit in fig.~\ref{siginv}
the momentum distribution at two different centre-of-mass angles.
Both distributions are fitted by a thermal model 
with angular dependent temperature parameters.

In this respect it is interesting to observe that the different channels
provide very different angular distributions. The $\phi$ mesons
which are created in BB collisions have a nearly isotropic
angular distribution. However, in the $\pi$B  and $\rho$B channels
the resulting angular distributions are significantly forward-backward
peaked. 
This anisotropy caused by the meson channels determines the total 
momentum distribution. This leads to the fact that the yield
in the backward lying CDC detector is larger than that in the 
HELITRON detector which is sensitive at midrapidity. 
This finding  seems to be in contradiction to 
the preliminary analysis by the FOPI group \cite{kottenew}. Also
results by Aichelin and Hartnack \cite{hartnack} 
within a quantum molecular dynamic transport code support
this behaviour of the different channels. As in their 
calculations the BB channels play a more important role they
obtain a nearly thermalised
$\phi$ distribution though the cross section 
is essentially smaller there.

\begin{figure}
\begin{center}
\includegraphics[width=8.6cm]{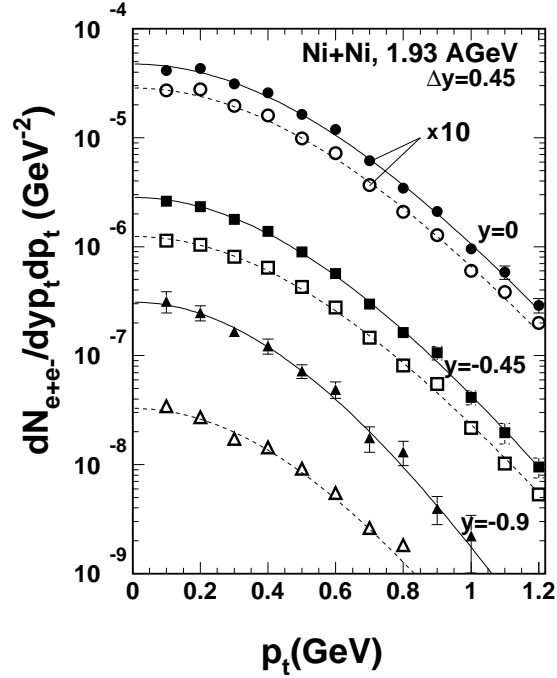}
\end{center}
\caption{ Transverse momentum spectra of electron-positron pairs
from $\phi$$\to$e$^+$e$^-$ decays  
at three different centre-of-mass rapidities $y$ shown by full symbols.
Corresponding open symbols 
indicate those electron-positron pairs which can be 
observed within the geometry
of the HADES detector. The efficiency of the 
electron and positron detection is not accounted for. The lines
are drawn to guide the eye.} 
\label{pt}
\end{figure}

The anisotropy can be studied by dedicated measurements with the HADES
detector.  To get an idea of the expected spectral shapes we display
in Fig.~\ref{pt} the transverse momentum distributions of the 
electron-positron pairs which stem from $\phi$ meson decays
for three rapidity slices. The open symbols in Fig.~\ref{pt} 
indicate the pairs which enter the HADES detectors\footnote{We thank 
R. Kotte for providing us with the routines of the 
HADES acceptance.}. 
One observes that the filtered e$^+$e$^-$ spectra resemble 
much the original  $\phi$ $\to$ e$^+$e$^-$ spectra depicted by the full
symbols in Fig.~\ref{pt}.
This suggests that the original $\phi$ distribution can be recovered
by the detector.

\section{Summary and discussion} 

In summary we have presented an analysis of $\phi$ meson production in
central heavy-ion collisions at beam 
energies of 1.69 and 1.93 A$\cdot$GeV.
As new channels we have included $\rho$B $\to \phi$N, 
$\pi\rho \to \phi$ and 
$\pi$ N$^* \to \phi$N and copious $\rho$ production via the resonance
decays $B^* \to \rho$N. Despite of enlarging the $\phi$
multiplicities, as follows within a transport code of BUU type,
the experimentally found values \cite{kotte,kottenew} are still
underestimated. 
This conclusion is not affected by the theoretical uncertainties
regarding the $\pi$ + N(1520) channel.
A momentum dependent soft mean field
for baryons and a weak mean field for the $\phi$ mesons have been
employed. We have not considered a possible change of the 
decay width of the $\phi$ meson as the momentum dependence of
the in-medium properties are poorly known at present.

With respect to the potential importance of the $\phi$ meson for the 
$K^-$ channel a dedicated measurement with enlarged statistics
would be highly welcome to resolve the $\phi$ puzzle.
It should be emphasised that the $\phi$ meson becomes now also 
accessible via the
$e^+ e^-$ decay channel in HADES measurements \cite{HADES}. 
The branching ratio 
between the electromagnetic and the hadronic decay will be
sensitive to the in-medium properties of mesons. Thus,
the $K^+ K^-$ decay channel should be addressed in future
experiments with both the up-graded FOPI apparatus and the HADES
detector due to the precise tracking system.

\subsection*{Acknowledgments} 

We thank R. Kotte for continuous information on his analyses of the
FOPI data. 
We are grateful to J.\ Aichelin and C.\ Hartnack for informing us
on their own analyses of the data. 
This work was supported in part by the German BMBF 
grant 06DR921, the DAAD scientific exchange program with Hungary,
and the National Fund for Scientific
Research of Hungary, OTKA T30171, T30855, and T32038.

\end{document}